\documentstyle[12pt,epsf]{article} 

\def\lsim{\mathrel{\lower2.5pt\vbox{\lineskip=0pt\baselineskip=0pt 
           \hbox{$<$}\hbox{$\sim$}}}} 
\def\gsim{\mathrel{\lower2.5pt\vbox{\lineskip=0pt\baselineskip=0pt 
           \hbox{$>$}\hbox{$\sim$}}}}

\begin{document}

\begin{flushright}
KGKU-00-01 \\ hep-ph/0001323
\end{flushright}

\vspace{10mm}

\begin{center}
{\large \bf Mediation of Supersymmetry Breaking \\
via Anti-Generation Fields}

\vspace{20mm}

Masato ITO$^a$ 
            \footnote{E-mail address: mito@eken.phys.nagoya-u.ac.jp}
and Takeo MATSUOKA$^b$ 
            \footnote{E-mail address: matsuoka@kogakkan-u.ac.jp}
\end{center}

\begin{center}
{
\it 
{}$^a$Department of Physics, Nagoya University, Nagoya, 
JAPAN 464-8602 \\
{}$^b$Kogakkan University, Nabari, JAPAN 518-0498 
}
\end{center}

\vspace{25mm}

\begin{abstract}
In the context of the weakly coupled heterotic string, 
we propose a new model of mediating supersymmetry breaking. 
The breakdown of supersymmetry in the hidden sector is transmitted 
to anti-generation fields via gravitational interactions. 
Subsequent transmission of the breaking to the MSSM sector 
occurs via gauge interactions. 
It is shown that the mass spectra of superparticles are 
phenomenologically viable.
\end{abstract}

\newpage 

The dynamical mechanism of supersymmetry(SUSY) breaking and its 
mediation to our visible world form the major unsolved problems of 
particle physics. 
As for the former, many models have been proposed\cite{DSB}. 
For phenomenological implications it is of more importance to 
clarify how SUSY breaking is communicated to the visible sector. 
Two ways of transmitting SUSY breaking to the visible sector 
have been proposed. 
The first is that gravitational interactions in $N=1$ supergravity 
theory play this role\cite{SUGRA}. 
The second is that gauge interactions play the role of the
messenger\cite{GMSB}. 
In the framework of superstring theory it seems that 
the gravity-mediated SUSY breaking scenario is plausible. 
In the context of the weakly coupled heterotic 
$E_{8}\times E^{\prime}_{8}$ superstring, 
$E^{\prime}_{8}$ is considered to be the hidden sector gauge symmetry. 
If supersymmetric gauge dynamics in the hidden sector, 
such as gaugino condensation, cause SUSY breaking to occur, 
the effects of the SUSY breaking will be transmitted to the 
observable sector through gravitational interactions. 
In these types of simple models\cite{FTdomi}, 
the $F$-components of moduli fields other than the dilaton are expected 
to develop their vacuum expectation values (VEVs). 
The gauginos acquire their soft SUSY breaking masses 
at one loop, but not at the tree level. 
As a consequence, the gaugino masses turn out to be too 
small compared to the soft scalar masses. 
Namely, we have the relations 
\begin{equation}
      m_{\lambda} \ll m_{0} \lsim m_{3/2} \,, 
\label{eqn1}
\end{equation}
where $m_{\lambda}$, $m_{0}$ and $m_{3/2}$ are the gaugino mass, 
the soft scalar mass and the gravitino mass, respectively. 
This mass spectrum of superparticles is phenomenologically 
unacceptable. 
Although it has been pointed out that this problem can be solved 
in strongly coupled heterotic $E_{8}\times E^{\prime}_{8}$ 
superstring theory (M-theory)\cite{H-W}\cite{problem}, 
there remains another problem regarding non-universality of the soft 
breaking mass parameters. 
Since in string theory the moduli fields are generally 
charged under the flavor symmetry, 
the soft scalar masses become flavor-dependent. 
As a result of these non-universal properties for the scalar masses, 
we are confronted with the so-called FCNC problem. 
With regard to this point, the gauge-mediated SUSY breaking scenario 
has the advantage of producing the universal soft scalar masses and 
then ensuring sufficient suppression of the FCNC. 

In this paper we propose a new model based on the weakly coupled 
heterotic string on the Calabi-Yau compactification, 
whose superparticle mass spectrum is phenomenologically 
viable and which is free from the FCNC problem. 
The breakdown of SUSY coming from the hidden sector gaugino condensation 
begins by being transmitted to anti-generation fields in the observable 
sector via gravitational interactions. 
This is due to the assumption that only VEVs of the K\"ahler 
class moduli fields $T$ acquire nonvanishing $F$-components. 
Subsequent transmission of SUSY breaking to the low-energy 
MSSM sector occurs via gauge interactions. 
Then the gaugino masses are of the same order as the soft scalar masses. 
In addition, due to the flavor-blind nature of gauge interactions, 
we obtain the universal soft scalar masses. 

The four-dimensional effective theory in the observable sector 
from the weakly coupled  Calabi-Yau string is characterized by 
$N=1$ SUSY, the $E_{6}$ gauge group, and massless matter fields
which belong to $\bf 27$ and $\bf 27^{\ast}$ representations in $E_{6}$. 
The massless chiral superfields apart from $E_{6}$ singlets 
consist of
\begin{equation}
   N_{f}\,\Phi({\bf 27})\; + 
     \;\delta\,(\Phi({\bf 27}) + {\overline \Phi}({\bf 27}^{\ast})) \,, 
\label{eqn2}
\end{equation}
where $N_{f}$ denotes the family number at low energies. 
$\delta$ sets of vector-like multiplets are included in 
the massless sector. 
The numbers $N_f + \delta$ and $\delta$ represent the generation number and 
the anti-generation number, respectively. 
These numbers coincide with the Hodge numbers $h^{21}$ and 
$h^{11}$ for the compactified manifold, respectively. 
We will assume $N_f = 3$ and $\delta = 1$ for the sake of simplicity. 
The particles beyond the MSSM are contained in ${\bf 27}$. 
Namely, in ${\bf 27}$ we have color-triplet Higgses $g$ and $g^{c}$ and 
a singlet $S$ as well as the quark superfield $Q=(U,D),U^{c},D^{c},$ 
lepton superfield $L=(N,E),N^{c},E^{c},$ and 
Higgs doublets $H_{u},H_{d}$. 
Moreover, there appear a dilaton field $D$, 
$h^{11}$ K\"ahler class moduli fields $T_{i}$, 
and $h^{21}$ complex structure moduli fields $U_{i}$. 
The VEV of a dilaton field $D$ determines the gauge 
coupling constant, and the VEVs of the moduli fields $U_{i}$ and $T_{i}$ 
parametrize the size and shapes of the compactified manifold. 
In the observable sector from the Calabi-Yau string, 
the superpotential $W$ contains trilinear terms 
in $\Phi $ and ${\overline \Phi}$ :
\begin{equation}
    W = h(U) \, \Phi^{3} + f(T) \, \overline{\Phi}^{3} \,.
\label{eqn3}
\end{equation}
Here $h(U)$ and $f(T)$ are the appropriate holomorphic functions of 
the moduli fields $U$ and $T$, respectively. 
In Eq. (\ref{eqn3}) we have omitted the generation indices. 
In Ref.$\cite{worldsheet}$ it is shown that 
the cubic terms of ${\overline \Phi}$ 
in the superpotential are corrected by instantons 
on the string worldsheet. 
It should be noted that in the Calabi-Yau string, 
$\Phi$ and ${\overline \Phi}$ 
couple separately to the $U$ moduli and the $T$ moduli, respectively
\cite{27}. 
As mentioned above, in the observable sector, the seed of SUSY breaking 
is expected to be a nonvanishing VEV of $F^{T}$
\cite{FTdomi}, 
which is the $F$-component of the moduli field $T$. 
The $F$-components of the dilaton $D$ and the moduli $U$
are assumed to have vanishing VEVs. 
This situation is the same as that in the moduli-dominated SUSY breaking 
scenario\cite{modulibreak}. 
Hereafter we denote the VEV of the moduli field $T$ by the same 
letter as the field. 
For example, we write 
$\langle T \rangle = T + \theta^{2}\,F^{T}$. 

In Eq. (\ref{eqn3}) the trilinear coupling of ${\overline \Phi}$ 
contains the terms 
\begin{equation}
    f(T)(\overline{S}\overline{H}_{u}\overline{H}_{d} + 
            \overline{S}\overline{g^{c}}\overline{g})\,.
\label{eqn4}
\end{equation}
If the singlet field $\overline{S}$ acquires a nonvanishing VEV 
$\langle \overline{S} \rangle$ \cite{E6}, 
then $\overline{H}_{u}$, $\overline{H}_{d}$, $\overline{g^{c}}$ 
and $\overline{g}$ in ${\overline \Phi}$ gain supersymmetric masses 
\begin{equation}
    m_{\overline \Phi_{1/2}} \sim f(T) \, \langle \overline{S} \rangle \,.
\label{eqn5}
\end{equation}
Also, we have the soft SUSY breaking scalar masses, 
which are given by 
\begin{equation}
   m_{\overline \Phi_{0}}^2 \sim f^{\prime}(T) 
                          F^{T}\langle \overline{S} \rangle \,, 
\label{eqn6}
\end{equation}
where $f^{\prime}(T)$ denotes the derivative with respect to $T$.
Therefore, the scalar particles in $\overline{H}_{u}$, $\overline{H}_{d}$, 
$\overline{g^{c}}$ and $\overline{g}$ have the masses squared 
$m_{\overline \Phi_{1/2}}^2 \pm m_{\overline \Phi_{0}}^2$. 

Since only the anti-generation fields 
${\overline \Phi}$ couple to the $T$ moduli, 
the fields ${\overline \Phi}$ behave as messenger fields 
in the framework of the gauge-mediated SUSY breaking scenario.
It should be noted that the flavor dependence of $\langle T_{i} \rangle$ 
is not transmitted to the generation fields $\Phi$.
In the GUT-type models which accommodate the MSSM at low energies, 
$\langle \overline{S} \rangle$ and 
$\langle S \rangle$ are expected to be larger than $10^{16}$ GeV
\cite{E6}. 
In this case these messenger fields will become sufficiently heavy 
compared to the electroweak scale. 
The other extra fields belonging to one set of the vector-like multiplet 
also become massive near or slightly above the intermediate energy scale 
via nonrenormalizable interactions \cite{E6}. 
On the other hand, due to vanishing $F$-terms of the $U$ moduli, 
the generation fields $\Phi$ cannot behave as messenger fields. 
The messenger fields ${\overline \Phi}$ have 
the quantum number of the standard model gauge group. 
Therefore, integrating out the messenger sector gives rise to 
gaugino masses at one loop. 
Estimating one-loop diagrams, one finds that the gaugino masses induced are
\begin{equation}
    m_{\lambda_i} \sim \frac{g^{2}_{i}}{16\pi^{2}}\Lambda \,, 
\label{eqn7}
\end{equation}
where 
\begin{equation}
   \Lambda = \left| \frac {f^{\prime}(T)}{f(T)} \right| \,F^{T}\,, 
\label{eqn8}
\end{equation}
and the $g_{i}$s are the gauge coupling constants ($i=1,2,3$). 
The soft scalar masses in the low-energy MSSM sector, 
in which $N_f$ sets of the generation fields $\Phi({\bf 27})$ 
remain light, 
arise at leading order from two-loop diagrams. 
Messenger fields, gauge bosons and gauginos take part 
in the internal lines of the two-loop diagrams. 
Consequently, the soft scalar masses induced are of the same order 
as the gaugino masses :
\begin{equation}
    m_{0} \sim m_{\lambda_i}\,. 
\label{eqn9}
\end{equation}
Provided that $\Lambda$ is $ {\cal O}(100\;{\rm TeV})$, 
the mass spectrum of superparticles is consistent with that in the MSSM. 
As mentioned above, the significant features of this scenario 
are that there are sufficient degeneracies among squarks (sleptons) 
to ensure adequate suppression of FCNC 
and that no new $CP$ phases are induced in soft SUSY breaking 
parameters. 

As a consequence of the worldsheet instanton effects, 
$f(T)$ is of the form 
\begin{equation}
    f(T) \propto e^{-\lambda T}\,, 
\label{eqn10}
\end{equation}
where $\lambda$ is a constant. 
${\rm Re}\,T$ is related to the compactification radius 
$R$ as ${\rm Re}\,T \sim R^{2}$ in units of the string scale. 
The string scale $M_s$ is defined by $(\alpha ')^{-1/2}$, 
where $\alpha '$ is the string tension.  
From Eq. (\ref{eqn10}), $\Lambda$ in Eq. (\ref{eqn8}) becomes 
$\Lambda = \lambda \,F^{T}$. 
In view of the fact that ${\rm Im}T$ corresponds to an axion-like 
field, 
it is adequate for us to require the periodicity of $f(T)$ 
in units of the string scale. 
In this context we postulate that 
$\lambda = {\cal O}(2 \pi M_s^{-1})$. 
Thus, we have 
\begin{equation}
    m_{\lambda _i} \sim \frac {g_i^2}{8\pi} 
                        \times \frac {F^T}{M_s}\,. 
\label{eqn11}
\end{equation}
With the scale $F^T/M_s = {\cal O}(20\;{\rm TeV})$, 
we find that the gaugino masses are on the order of a few hundred GeV. 

Now we proceed to discuss the gravitino mass. 
In the weakly coupled heterotic string, the K\"ahler potential 
is given by \cite{kahler}
\begin{equation}
    K = - \ln (D+D^{\ast}) - 3\ln (U+U^{\ast})
                           - 3\ln (T+T^{\ast}) + \cdots . 
\label{eqn12}
\end{equation}
Here the dots represent the higher-order terms. 
Using the above K\"ahler potential and assuming 
a vanishing cosmological constant,
we obtain the gravitino mass
\begin{equation}
     m_{3/2} = \frac{F^{T}}{2{\rm Re}\,T}\sim \frac{F^{T}}{2M_s}\,.
\label{eqn13}
\end{equation}
In string theory, the compactification radius is expected to be 
around the inverse string scale. 
From Eqs. (\ref{eqn9}), (\ref{eqn11}) and (\ref{eqn13}) we obtain 
the phenomenologically viable relations 
\begin{equation}
     m_{0} \sim m_{\lambda_i} \sim \frac {m_{3/2}}{10}\,. 
\label{eqn14}
\end{equation}

Since the gaugino mass and the scalar mass are also induced by
gravitational interactions, 
we must study whether or not the gauge-mediated contribution 
is dominated by a gravity-mediated contribution. 
In the weakly coupled heterotic string, 
the gauge-kinetic function in the observable sector is of the form 
$f_6= D + \epsilon T$, and 
we have $D \gg \epsilon T$ in the weak coupling region. 
The second term in $f_6$ comes from the string one-loop correction 
and $\epsilon$ is a small constant fixed by gauge and gravitational
anomaly\cite{kahler}\cite{f6}. 
The gravity-mediated contribution to the gaugino mass is given by 
\begin{equation}
    m_{\lambda}^{\rm (gr)} = \sum_{m = D,T,U} \frac{F^{m}\partial_{m}f_{6}}
                            {2{\rm Re}\,f_{6}}
                  \sim \frac{\epsilon F^{T}}{2{\rm Re}\,D}\,. 
\label{eqn15}
\end{equation}
Since we have ${\rm Re}\,D = 4\pi/(g_i^2)$ at the unification 
scale, 
Eq. (\ref{eqn15}) cac be rewritten as 
\begin{equation}
    m_{\lambda_i}^{\rm (gr)} \sim 
           \frac{g_i^2}{8\pi} \times \frac {\epsilon F^T}{M_s}\,. 
\label{eqn16}
\end{equation}
From Eqs. (\ref{eqn11}) and (\ref{eqn16}) 
the ratio of $m_{\lambda_i}^{\rm (gr)}$ to $m_{\lambda_{i}}$ becomes
\begin{equation}
    \frac{m_{\lambda _i}^{\rm (gr)}}{m_{\lambda_i}}
                \sim \epsilon \ll 1\,. 
\label{eqn17}
\end{equation}
Further, the gravity-mediated contribution to the scalar mass
is given by
\begin{equation}
 m_{0}^{\rm (gr)}=\sqrt{\frac{\epsilon {\rm Re}T}{{\rm Re}D}}m_{3/2}\sim
             \frac{F^{T}}{M_{s}}\sqrt{\epsilon\frac{g^{2}_{i}}{16\pi}}
 \label{eqn18}
\end{equation}
up to a one-loop correction \cite{problem}.
Then the ratio of $m_{0}^{\rm (gr)}$ to $m_{0}$ becomes
\begin{equation}
 \frac{m_{0}^{\rm (gr)}}{m_{0}}\sim
 \sqrt{\epsilon\frac{4\pi}{g^{2}_{i}}}\sim 5\sqrt{\epsilon}\ll 1\,.
 \label{eqn19}
\end{equation}
where we have taken $g^{2}_{i}/(4\pi)\sim 1/25$ at the unification scale.
As can be deduced from Eqs. ($\ref{eqn17}$) and ($\ref{eqn19}$), 
the gauge-mediated contribution to the gaugino mass
and scalar mass are dominant compared to the gravity-mediated
contribution. 

\begin{figure}
      \epsfxsize=10cm
\centerline{\epsfbox{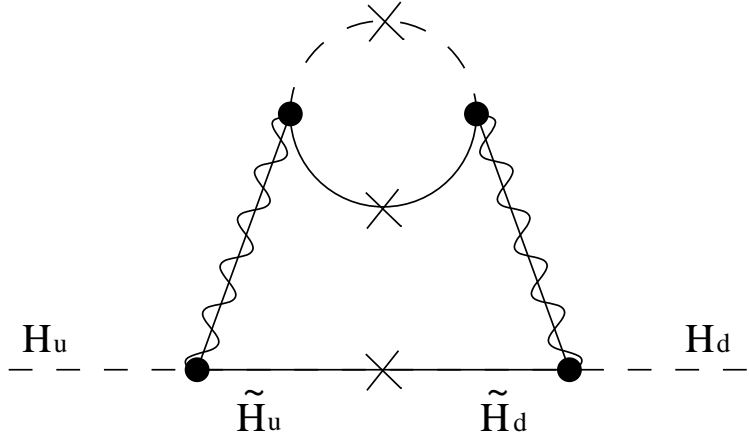}}
\caption{The $B\mu$ term is generated in the two-loop diagram.}
\label{fig1}
\end{figure}

Finally, we discuss the $\mu$ problem. 
In the present model the $\mu$ term comes from the first term in 
Eq. (\ref{eqn3}) with $\mu \sim h(U)\langle S \rangle $. 
The $B \mu $ term is generated in the two-loop diagram as shown in
Fig.\ref{fig1}. 
The messenger fields (${\overline \Phi}$) 
propagate in the innermost loop. 
The external lines are the scalar fields $H_{u}$ and $H_{d}$ and 
the remaining internal line has the fermion mass insertion 
$\mu H_{u}H_{d}$. 
The ratio of $B\mu$ to $\mu ^2$ becomes 
\begin{equation}
   \frac{B\mu}{\mu ^2} \sim 8 \left(\frac{g_{2}}{4\pi}\right)^{2}
                            \frac{m_{\lambda_{2}}}{\mu}\,
                      \ln\left(\frac{m_{\lambda_{2}}}{\mu}\right)\,
                    \sim {\cal O}(1)\,,
\label{eqn20}
\end{equation}
provided that $m_{\lambda_{2}}/\mu \sim {\cal O}(10)$. 
Introducing a certain type of flavor symmetry, 
we can obtain $\mu \sim h(U)\langle S \rangle\sim {\cal O}(100\;{\rm GeV})$
\cite{E6}. 
Thus, the $\mu$ problem is solved. 

In conclusion, we propose a new model of SUSY breaking mediation 
on the basis of the weakly coupled heterotic string on the Calabi-Yau 
compactification. 
In this model, SUSY breaking is mediated to the MSSM sector by 
two phases. 
In the first phase, SUSY breaking in the hidden sector is 
transmitted to anti-generation fields via gravitational 
interactions. 
Subsequent transmission of SUSY breaking occurs via 
gauge interactions. 
Although the moduli fields are charged under the flavor symmetries, 
due to the flavor-blind nature of gauge interactions, we obtain the
universal soft scalar masses.
The mass spectra of superparticles obtained here are
phenomenologically viable.

\section*{Acknowledgements}
\ \ \ A part of this work was done at the Department of Physics 
Engineering, Mie University, Tsu, Japan. 
The authors are grateful to Professor Y. Abe and Assistant Professor
M. Matsunaga for their hospitality. 
One of the authors (T. M.) is supported in part by 
a Grant-in-Aid for Scientific Research from the Ministry 
of Education, Science, Sports and Culture, Japan 
(No . 10640256). 



\end{document}